\documentclass[fleqn,11pt,twoside]{article}

\usepackage{amsthm,amsthm,amssymb, color, xcolor,epsfig, graphics, subfigure}

\usepackage{amsmath, graphicx, latexsym, lscape}
\usepackage{cite}
\usepackage{caption}
\usepackage{xcolor}
\usepackage{dutchcal}
\usepackage{float}
\usepackage{hyperref}

\makeatletter
\newcommand{\copyrightnote}[2]{{\renewcommand{\thefootnote}{}
 \footnotetext{\small\it
\begin{flushleft}
 \copyright \ #1   #2  
\end{flushleft}}}}

\newcommand{\Name}[1]{\begin{flushleft}
                       \LARGE \bf #1
                       \end{flushleft}\vspace{-3mm}}

\newcommand{\Author}[1]{\begin{flushleft}
                       \it #1 \end{flushleft}}

\newcommand{\Address}[1]{\begin{flushleft}
                       \it #1 \end{flushleft}}

\newcommand{\Date}[1]{\begin{flushleft}
                      \small  \it #1 \end{flushleft}}

\newcommand{\evenhead}{Author \ name}
\newcommand{\oddhead}{Article \ name}

\renewcommand{\@evenhead}{
\hspace*{-3pt}\raisebox{-15pt}[\headheight][0pt]{\vbox{\hbox to \textwidth
{\thepage \hfil \evenhead}\vskip4pt \hrule}}}
\renewcommand{\@oddhead}{
\hspace*{-3pt}\raisebox{-15pt}[\headheight][0pt]{\vbox{\hbox to \textwidth
{\oddhead \hfil \thepage}\vskip4pt\hrule}}}
\renewcommand{\@evenfoot}{}
\renewcommand{\@oddfoot}{}

\def\dblone{\hbox{$1\hskip -1.2pt\vrule depth 0pt height 1.6ex width 0.7pt \vrule depth 0pt height 0.3pt width 0.12em$}}
\long\def\@makecaption#1#2{%
  \vskip\abovecaptionskip
  \sbox\@tempboxa{\small \textbf{#1.}\ \ #2}%
  \ifdim \wd\@tempboxa >\hsize
    {\small \textbf{#1.}\ \ #2}\par
  \else
    \global \@minipagefalse
    \hb@xt@\hsize{\hfil\box\@tempboxa\hfil}%
  \fi
  \vskip\belowcaptionskip}

\newcommand{\JNMPnumberwithin}[3][\arabic]{%
  \@ifundefined{c@#2}{\@nocounterr{#2}}{%
    \@ifundefined{c@#3}{\@nocnterr{#3}}{%
      \@addtoreset{#2}{#3}%
      \@xp\xdef\csname the#2\endcsname{%
        \@xp\@nx\csname the#3\endcsname .\@nx#1{#2}}}}%
}

\newcommand{\resetfootnoterule} {
  \renewcommand\footnoterule{%
  \kern-3\p@
  \hrule\@width.4\columnwidth
  \kern2.6\p@}
}

\renewcommand{\footnoterule}{}

\makeatother

\theoremstyle{definition}

\begin{document}

\renewcommand{\evenhead}{ {\LARGE\textcolor{blue!10!black!40!green}{{\sf \ \ \ ]ocnmp[}}}\strut\hfill O Ragnisco and F Zullo}
\renewcommand{\oddhead}{ {\LARGE\textcolor{blue!10!black!40!green}{{\sf ]ocnmp[}}}\ \ \ \ \   
$N$-species Volterra system and
superintegrable Hamiltonian system
}

\thispagestyle{empty}
\newcommand{\FistPageHead}[3]{
\begin{flushleft}
\raisebox{8mm}[0pt][0pt]
{\footnotesize \sf
\parbox{150mm}{{Open Communications in Nonlinear Mathematical Physics}\ \  \ {\LARGE\textcolor{blue!10!black!40!green}{]ocnmp[}}
\ \ Vol.5 (2025) pp
#2\hfill {\sc #3}}}\vspace{-13mm}
\end{flushleft}}

\FistPageHead{1}{\pageref{firstpage}--\pageref{lastpage}}{ \ \ Article}

\strut\hfill

\strut\hfill

\copyrightnote{The author(s). Distributed under a Creative Commons Attribution 4.0 International License}

\Name{The N-species integrable Volterra system as a maximally superintegrable Hamiltonian system.}

\Author{O. Ragnisco$^a$, F.Zullo$^b$}

\Address{$^a$
Dipartimento di Matematica e Fisica, Universit\`a degli Studi ``Roma TRE" (retired)\\
Via della Vasca Navale 84,00146 Roma\\
e-mail: oragnisco@gmail.com\\
$^b$
DICATAM, Universit\`a degli Studi di Brescia
\\via Branze, 38 - 25123 Brescia, Italy \&\\
INFN, Milano Bicocca\\
Piazza della Scienza 3, Milano, 20126, Italy\\
e-mail: federico.zullo@unibs.it}

\Date{Received May 20, 2025; Accepted May 27, 2025}

\setcounter{equation}{0}

\begin{abstract}
\noindent 
The results presented in this paper are a natural development of those described in the paper {\it The Volterra Integrable case. Novel analytical and numerical results} (OCNMP Vol.4 (2024) pp 188-211), where the authors reconsidered the integrable case  of the Hamiltonian $N$-species Lotka-Volterra system, introduced by Vito Volterra in 1937. There, an alternative approach for constructing the  integrals of motion has been proposed, and compared with the old Volterra approach. Here we go beyond, and show that in fact the model introduced by Volterra and studied by us is not just integrable, but is maximally superintegrable and reducible to a system with only one degree of freedom regardless of the number of  species considered. We present both analytical and numerical results.
\end{abstract}

\label{firstpage}


\section{Introduction}
More than a century ago, independently, A.J.Lotka and V. Volterra \cite{Lotka1},\cite{Lotka2} \cite{Volterra} elaborated a simple but quite effective model able to describe the evolution of a two species biological system, the so-called predator-prey model. Later on, V.Volterra realized that
this model was just the simplest example in a large class of biological, or rather ecological systems with pairwise
interaction. In fact, in the late thirties of the past century  he constructed a dynamical system modelling the pairwise interaction of N-species \cite{Volterra},\cite{Volterra1937},  whose behaviour fulfilled  what he called the three fundamental laws of biological fluctuations (see \cite{Volterra1937}, pp. 20-21). He was working in the  framework of conservative models, which certainly are  not the only possible generalization of the original predator-prey system. Indeed,  the current literature is  quite rich of papers dealing both with conservative and dissipative models, see for instance \cite{fernandez} - \cite{Van der Kamp}.

The main ingredients of the system are the natural growth rate $\epsilon_j$ of each species and an $N\times N$ matrix $A_{j,k}$ describing the interaction between the species $j$ and the species $k$. Being interested in conservative models, he postulated the matrix $A$ to be skew-symmetrizable  and in fact he assumed it to be skew-symmetric. He found the Lagrangian and Hamiltonian formulation of his system, essentially by duplicating the number of variables through the introduction of what he called the ``quantity of life'', i.e. the time-integrals of the numerosities of the species. Further he distinguished between even and odd number of species: in the odd case, the interaction matrix $A$, being  skew symmetric, is singular and consequently the equilibrium configuration, whenever it exists, is not unique. Volterra concluded that  for non-zero growth coefficients the equilibrium states would have been impossible, and the number of individuals in some species will have grown  indefinitely or go to zero, so that on the long run only an even number of species would have survived. As we pointed out in \cite{SRTZ}, curiously enough, Volterra did not take into account the fact that complete integrability could change, even drastically, the above scenario. So, the aim we pursue and in fact attain in the present paper is the  characterization of the integrable Volterra  system discovered in \cite{Volterra1937} and studied in \cite{SRTZ} as a Hamiltonian {\emph{maximally superintegrable}} system.

Accordingly, in Section (\ref{sec2}), we start by recalling briefly the integrable version of the N-species Volterra system, observing  that the image of the matrix $A$ (hereafter denoted as ${\mathcal I}$) is a 2-dimensional subspace of the  N-dimensional Euclidean space ${\mathbb E}_N$, while its kernel, hereafter denoted as ${\mathcal K}$,  is (obviously) $(N-2)-$dimensional. 
 Then, we parametrize the two subspaces  ${\mathcal  I}$ and  ${\mathcal K}$, decompose a generic vector in its ${\mathcal K}$ and ${\mathcal I}$ components, and choose the coordinate system on ${\mathcal  I}$  as suggested by the structure of the matrix  $A$. We write down the evolution equations for the $N$-species Vollterra system in these new coordinates, which turn out to be canonical ones, and suitable to describe the system as a Hamiltonian system with one degree of freedom. For the sake of completeness we consider in some detail the $N=2$ case, that exhibits non-trivial differences with respect to any $N>2$ case. Finally, we  comment about the existence of equilibrium configurations, whose collection we denote by ${\mathcal E}$, and establish a connection between ${\mathcal E}$ and ${\mathcal K}$.
 
In Subsection (\ref{sec2.1}) we present an alternative approach, algebraic in nature, where analogous results are derived  starting from the spectral decomposition of the matrix $A$.

In Section (\ref{sec3}) we  comment about the results contained in \cite{SRTZ}, and discuss the fate of the complete integrability structure find out there, concluding that the system under scrutiny is in fact {\emph{maximally superintegrable}}(see for instance \cite{Evans} - \cite{Tremblay}).

In Section (\ref{seclin}), in order to get a qualitative insight in the long-time behavior of our  system, we look at its linearized version, identifying  sufficient conditions for the existence of bounded  orbits. Also, we remember a sufficient condition for the existence of bounded orbits in the fully non-linear case given in \cite{SRTZ}. 
In Section (\ref{sec5}) we show a number of examples of orbits and  trajectories  of the system written in terms of the canonical variables, corresponding to different choices of the parameters  and do  the comparison with the corresponding results displayed in \cite{SRTZ}.

In Section (\ref{sec6}) we   raise some comments on the relation between the integrability structure presented here, the one contained in \cite{SRTZ} and the original one proposed by Volterra in \cite{Volterra1937}. In the end  we  mention some possibly interesting  developments of the present research.

\section{The integrable Volterra system revised}\label{sec2}

\noindent
The equations for the $N$-species Volterra system read \cite{Volterra1937}
\begin{equation}
\frac{d N_r}{dt} = \epsilon_r N_r + \sum_{s\ne r=1}^N A_{rs}N_rN_s ~~(r=1,\cdots,N)\label{model}.
\end{equation}
In (\ref{model}), we have set all the parameters  introduced in \cite{Volterra1937} $\beta_r =1 ~\forall r$; $\epsilon_r$ are the natural growth coefficients of each species and $A_{rs} $ are interaction coefficients between species $r$ and species $s$ that account for the effects 
 of encountering between two individuals. In the integrable $N-$species case, the $N\times N$ interaction matrix $A$ has the form: 
\begin{equation}
A_{rs}=\epsilon_r\epsilon_s(B_r-B_s)\label{matrix}
\end{equation}
\noindent
where all the $ \epsilon_k$ and the $B_k$ are real non-zero numbers, and moreover $B_j\ne B_k$ if $j\ne k$.  Formula (\ref{matrix}) means that we can write 
\begin{equation}
A=[B, \epsilon\otimes \epsilon] \equiv (B \epsilon)\otimes \epsilon-\epsilon\otimes (\epsilon B), \label{matrixform}
\end{equation}
where $\epsilon$ is the vector of components $\epsilon_j, (j=1,\dots,N)$, and $B=diag(B_1,\dots,B_N)$, implying that $A$ has rank $2$.

\noindent
 So,  ${\mathcal I}$ is $2$-dimensional and  ${\mathcal K}$ is $(N-2)$-dimensional.
Moreover ${\mathcal I}$ and ${\mathcal K}$ are two orthogonal subspaces with respect to the inner product $(u,v)\equiv \sum_j u_j v_j$ and the $N$-dimensional Euclidean space ${\mathbb E}_N$ can be written as the 
direct sum of them:

\[
{\mathbb E}_N = {\mathcal I} \oplus {\mathcal K}.
\]
We remark that the formula (\ref{matrixform})
besides telling us that $A$ is a matrix of rank 2, defines as well its image as a linear  operator on ${\mathbb E}_N$: ${\mathcal I}$ is the linear span of two vectors, which are linearly independent because the coefficients $B_j$ are all distinct, namely $ \epsilon$ and $\eta$, of components $\epsilon_j$ and $\eta_j=B_j\epsilon_j$. It follows that the matrix elements of $A$ are given by $A_{jk}= \epsilon_k\eta_j-\eta_k\epsilon_j$.  ${\mathcal K}$ is naturally defined as the $(N-2)$-dimensional subspace of 
$\textrm{E}_N$ orthogonal to ${\mathcal I}$, i.e. such that its generic vector $\textrm{X}$ fulfils the linear  conditions 

\begin{equation}\label{kerA}
(\textrm{X},\epsilon)=(\textrm{X},\eta)=0.
\end{equation}

\noindent
The components of a  generic vector $\textrm{X}$  $\in {\mathcal K}$ will be denoted as $x_j$, and a coordinate realization will be given in ( \ref{coordX}).
On the other hand, the affine variety of equilibrium configurations  ${\mathcal E}$ is the  set  of elements  $\textrm{Z}$ fulfilling
\begin{equation}\label{Z}
(\textrm{Z},\epsilon)=0,~ (\textrm{Z},\eta)=1.
\end{equation}
 
 \noindent
Their components will  be denoted as $z_j$, and  an explicit realization in coordinates will be given in (\ref{coordZ}).

\noindent
It is convenient to rewrite the evolution equations (\ref{model}) In terms of the so-called logarithmic variables defined by $y_j := \log N_j$, getting:
\begin{equation}
\dot y_j= \epsilon_j +\sum_{k=1}^N A_{jk}\exp(y_k) = \epsilon_j +( \eta_j\sum_{k=1}^N\epsilon_k-\epsilon_j\sum_{k=1}^N\eta_k)\exp(y_k)\label{llog}
\end{equation}

\noindent
We represent any vector  $\in {\mathbb E}_N$ as a linear combination of a vector $ \in {\mathcal I}$ and a vector $\in {\mathcal K}$ as follows:
\begin{equation}
y_j=P\epsilon_j+Q\eta_j+ \sum_{k=1}^{N-2}R_k \tau^{(k)}_j\label{decompoy}
\end{equation}
\noindent
In (\ref{decompoy}) the expression  $\sum_{k=1}^{N-2}R_k \tau^{(k)}_j$ represents the decomposition of a generic element $\in {\mathcal K}$ along a given orthonormal basis.
\noindent
 The evolution equations (6) take the form:
\begin{equation}
\dot P \epsilon_j+\dot Q \eta_j + \sum_{k=1}^{N-2} \dot R_k \tau^{(k)}_j= \epsilon_j+\eta_j\sum_{k=1}^N \epsilon_k\exp(y_k) -\epsilon_j\sum_{k=1}^N \eta_k\exp(y_k)
\end{equation}
\noindent
entailing:
\begin{eqnarray}
\dot P= 1-\sum_{k=1}^N\eta_k \exp(y_k),\label{eveq2a}\\
\dot Q = \sum_{k=1}^N\epsilon_k \exp(y_k), \label{eveq2b}\\
\dot R_k =0, \quad k=1...N-2\label{eveq2c}
\end{eqnarray}

\noindent
But also one has:
\[
 \exp(y_k)=\exp(P\epsilon_k+Q\eta_k + \sum_{j=1}^{N-2} R_j\tau^{(j)}_k) \equiv \exp(P\epsilon_k+Q\eta_k){\mathcal C_k},
 \]
 where we defined $\mathcal{C}_k=\exp(\sum_{j=1}^{N-2} R_j\tau^{(j)}_k)$. We can  rewrite the equations  (\ref{eveq2a}, \ref{eveq2b})  as:
\begin{eqnarray}
\dot P= 1-\sum_{k=1}^N{\mathcal C}_k\eta_k\exp(P\epsilon_k+Q\eta_k),\label{eveq3a}\\
\dot Q = \sum_{k=1}^N{\mathcal C}_k\epsilon_k\exp(P\epsilon_k+Q\eta_k).\label{eveq3b}
\end{eqnarray}

\noindent
It turns out that the Hamiltonian structure of the above equations is quite simple, being the standard canonical one. 
Also, the $N-2$ constants of motion appearing in equation (\ref{eveq2c}) are not just conserved quantities, but a set of Casimirs\footnote{We remember that, given a Poisson manifold $M$, a Casimir related to the corresponding Poisson structure is any nonconstant function on $M$ that Poisson commutes with any function on $M$, i.e. if $P$ is the Poisson tensor, the Casimir is a function whose differential belongs to the Kernel of $P$.}  for the Poisson structure related  to (\ref{llog}). 
\noindent
Let us remind  \cite{SRTZ} that the Hamiltonian formulation reads:
\[
\dot y_j= \{y_j, {\mathcal H}\}
\]
\noindent
where the Poisson bracket between two smooth functions on ${\mathbb  E}_N$, say ${\mathcal F}, {\mathcal  G}$, is defined as:
\begin{equation}
{\{\mathcal F},{\mathcal  G}\}:=\sum_{j,k=1}^N\frac{\partial {\mathcal F}}{\partial y_j}A_{jk}\frac{\partial {\mathcal G}}{\partial y_k}\label{PB}
\end{equation}
The previous Poisson bracket is not kernel-free and  indeed we have a set of $N-2$ Casimirs: if we consider any vector  $\textrm{X}$ belonging to ${\mathcal K}$, whose components $x_k$  obey the relations (\ref{kerA}), then the Casimirs are explicitly given by (see also \cite{CDE}): 
\begin{equation}
\sum_{n=1}^N y_n\sum_{k=1}^{N-2} R_k\tau^{(k)}_n=\sum_{k=1}^{N-2}R_k(Y,\tau^{(k)})= (Y,X)\label{Cas}
\end{equation}
\noindent
Formula $(\ref{Cas})$ is easily proved: for an arbitrary function $F$ one has
\begin{equation}
\{F,(Y,X)\}= \sum_{j,k=1}^N{\frac{\partial F}{\partial y_j}}A_{jk}{\frac{\partial (Y,X)}{\partial y_k}}=\sum_{j,k=1}^N{\frac{\partial F}{\partial y_j} A_{jk}{x_k}}=0.\label{PC}
\end{equation}
The functions $R_k$ are explicitly given by $(Y,\tau^k)$, i.e. is the projection of the dynamical variables on the $k$-th orthonormal base vector spanning the Kernel. These functions can be taken as a basis for the set of $N-2$ Casimirs. The functions $\mathcal{C}_k=\exp(\sum_{j=1}^{N-2} R_j\tau^{(j)}_k)$ are then an exponential of a linear combination of Casimirs, whose coefficients $\tau^{(j)}_k$ depend only on the parameters of the models and are constant as well.

\noindent
From (\ref{PB}) it follows:
\begin{eqnarray}
\dot P=-\sum_{k=1}^N\eta_k\frac{\partial {\mathcal H}}{\partial y_k}\\
\dot Q = \sum_{k=1}^N\epsilon_k\frac{\partial {\mathcal H}}{\partial y_k}.
\end{eqnarray}

\noindent
Taking into account the formulas:

\[\frac{\partial}{\partial P}=\sum_{k=1}^N\epsilon_k\frac{\partial}{\partial y_k}\]
\[\frac{\partial}{\partial Q}=\sum_{k=1}^N\eta_k\frac{\partial}{\partial y_k}\]

\noindent
we can rewrite the Hamilton  equations in the canonical form:

\begin{equation}
\dot P= -\frac{\partial {\mathcal H}}{\partial Q}; ~~\dot Q = \frac{\partial {\mathcal H}}{\partial P}
\end{equation}

\noindent
and the Hamiltonian reads:

\begin{equation}\label{h1}
{\mathcal H}=\sum_{k=1}^N \mathcal{C}_k\exp(P\epsilon_k+Q\eta_k) -Q
\end{equation}

Let us write the relations linking the coordinates $(P,Q)$ with the vector $Y$ with components $y_j$. From the relations (\ref{kerA}) and (\ref{Z}) we get:
\begin{equation}
Q=\frac{(Y, A\epsilon)}{(\eta,A\epsilon)}, \quad P=- \frac{(Y, A\eta)}{(\epsilon,A\eta)},
\end{equation}
\noindent
where $(\epsilon,A\eta)=(\epsilon,\eta)^2-|\epsilon|^2|\eta|^2$ is strictly less than $0$ due to  Cauchy-Schwartz inequality (see for instance \cite{Kreyszig}), since the vectors $\epsilon$ and $\eta$ are not parallel.
 We notice  that, from equation (\ref{decompoy}), the coordinate $Q$ can be also written as:
\begin{equation}
Q=(Y,Z)-(X,Z)\label{Q}
\end{equation}

\noindent
i.e. $Q$ is the projection of the component of $Y$ belonging to  ${\mathcal I}$ on the space $\mathcal{E}$ of equilibrium configurations. The Hamiltonian (\ref{h1}) depends both on the Casimirs and the equilibrium configurations: this is  indeed well known, see e.g. \cite{SRTZ} and references therein. Further, the term $(X,Z)$ can be omitted in the Hamiltonian since it is just a constant term not containing the dynamical variables.

\medskip

\noindent
It is worthwhile to notice that the $N=2$ case, namely the original Lotka-Volterra system \cite{Volterra}, is peculiar. We discuss just the system in the coordinates $(P,Q)$.

\begin{enumerate}
\item
First of all,  only in this case $ {\mathcal K}$ is an empty set, while ${\mathcal E}$  consists of a\ single point. Indeed, from (\ref{kerA}) and (\ref{Z}) we get:
\begin{eqnarray} 
\epsilon_1x_1+\epsilon_2x_2=0\label{ker1}\\
\eta_1x_1 +\eta_2x_2=0\label{ker2}
\end{eqnarray}

\noindent
which has no nonzero solutions, since $\Delta :=\epsilon_1\eta_2-\epsilon_1\eta_2$ is $\ne 0$,
and
\begin{eqnarray}
\epsilon_1 z_1 +\epsilon_2 z_2=0\label{2eq1}\\
\eta_1z_1+\eta_2z_2=1\label{2eq2}
\end{eqnarray}

\noindent
yielding the equilibrium position:

\[(z_1^, z_2)=(\frac{\epsilon_2}{\Delta},-\frac{\epsilon_1}{\Delta})\]
\noindent
implying that, assuming for instance $\Delta >0$ , it belongs to the first quadrant iff $\epsilon_1<0,~\epsilon_2>0$.
\item
Moreover, for $N=2$ we have ${\mathcal C}_k=1 (k=1,2)$. So, the equations of motion read:

\[\dot P= 1-\sum_{k=1}^2 \eta_k\exp(P\epsilon_k+Q\eta_k)\]
\[\dot Q = \sum_{k=1}^2\epsilon_k\exp(P\epsilon_k+Q\eta_k)\]
and clearly constitute a {\it closed} nonlinear differential system for the unknowns $P,Q$.

\noindent
By introducing the new variables:
\begin{equation}
S_j=\epsilon_jP+\eta_jQ~(j=1,2)\label{S}
\end{equation}

\noindent
it follows:
\begin{eqnarray}
P=\frac{\eta_1S_1-\eta_2 S_2}{\epsilon_1\eta_2-\epsilon_2\eta_1},\label{P}\\
Q=\frac{\epsilon_1S_1-\epsilon_2S_2}{\epsilon_1\eta_2-\epsilon_2\eta_1}.\label{Q}
\end{eqnarray}

\noindent
The equations of motion are rewritten as  

\begin{eqnarray}
\dot S_1=\epsilon_1-\Delta \exp(S_2)=-\Delta \frac{\partial{\mathcal H}}{\partial S_2}\label{finr1dot}\\
\dot S_2= \epsilon_2 +\Delta \exp(S_1)= \Delta  \frac{\partial {\mathcal H}}{\partial S_1}\label{finr2dot}
\end{eqnarray}

\noindent
where the Hamiltonian is  given by:
\begin{equation}
{\mathcal H}=\sum_{k=1}^2 \exp(S_k) +\frac{\epsilon_2S_1-\epsilon_1S_2}{\Delta},
\end{equation}

\noindent
Of course, the Hamiltonian can be found directly by integrating the orbit equation:
\begin{equation}
dS_1(\epsilon_2+\Delta \exp(S_1)=dS_2(\epsilon_1-\Delta \exp(S_2)\label{orbit}
\end{equation}
\end{enumerate}

\medskip

\medskip
We end this section by presenting an explicit characterization of the sets ${\mathcal K}$ and ${\mathcal  E}$.

\noindent
If one chooses two different arbitrary indices among $1,\ldots,N$, say $m$ and $n$, then a coordinate realization of the vectors $\textrm{X}$ (\ref{kerA}) can be $\textrm{X}=(x_1,\ldots,x_N)$, where 
\begin{equation}
x_m=\sum_{k\neq m}{\frac{x_k\epsilon_k (B_n-B_k)}{\epsilon_m(B_m-B_n)}}, \quad x_n=\sum_{k\neq n}{\frac{x_k\epsilon_k (B_m-B_k)}{\epsilon_n(B_n-B_m)}}, \quad m\neq n\label{coordX}
\end{equation}

\noindent
Also, a coordinate realization of the elements $\textrm{Z}$ $\in {\mathcal E}$ is provided by:
\begin{equation}\begin{split}
&z_m=\sum_{k\neq m}{\frac{z_k\epsilon_k (B_n-B_k)}{\epsilon_m(B_m-B_n)}}+\frac{1}{\epsilon_m(B_m-B_n)}, \\ 
&z_n=\sum_{k\neq n}{\frac{z_k\epsilon_k (B_m-B_k)}{\epsilon_n(B_n-B_m)}}+\frac{1}{\epsilon_n(B_n-B_m)}, \quad m\neq n\label{coordZ}
\end{split}\end{equation}
where again $m$ and $n$ are two arbitrary indices $1,\ldots,N$. The choices of $x_k$ and $z_k$, for $k \neq m,n$, are arbitrary. For example, in the $N=3$ case, one can take $m=1$ and $n=2$, so that  the only remaining index is $k=3$, i.e. $z_3$ is arbitrary.

\bigskip
\subsection{An (alternative) algebraic approach relying on the spectral decomposition, leading to an analogous Hamiltonian formulation.}\label{sec2.1}

\noindent 
The matrix $A$, being real and skew-symmetric (which ensures that $iA$ is hermitian), is amenable to a spectral decomposition:
\begin{equation}
A= i\omega( v^+\otimes \bar v^+- v^-\otimes \bar v^-),\label{spectrdec}
\end{equation}
\noindent
where $v^+$ and $v^-$ are the mutually orthogonal eigenvectors corresponding to the eigenvalues $\pm i\omega\in i{\mathbb  R}$ and the overline denotes complex conjugation.
The two eigenvectors are orthogonal with respect to the complex scalar product: throughout this subsection we set $(u,v)= \sum_{k=1}^N\bar u_k v_k$.

\noindent
Let us remind the definition of vectors $\epsilon$ and $\eta$:
\begin{equation}
\epsilon :=\{\epsilon_j\}_{j=1,\dots N}; \;~\eta:=\{B_j\epsilon_j\}_{j=1,\dots N}\label{basis}
\end{equation}
\noindent
so that ${\mathcal K}$ is given by (\ref{kerA}), and in coordinates by (\ref{coordX}); on the other  hand, the set of equilibrium configurations is given  by (\ref{Z}), and in coordinates by (\ref{coordZ}).
\noindent

 \noindent
We notice that from the eigenvalue equation
\begin{equation}\label{eigeq}
\epsilon_j\sum_{k=1}^N\epsilon_k(B_j-B_k)v_k= \lambda v_j
\end{equation}
one has  $v^-_j=\bar v^+_j,~~v^-_j=\bar v^+_j$ and we can  rewrite (\ref{spectrdec}) in the form:
 \begin{equation}
 A_{jk}=i\omega(v^+_jv^-_k - v^-_jv^+_k)\rightarrow A= i\omega(v^+\otimes v^- -v^-\otimes v^+)\label{Askew}
 \end{equation}
which makes its skew-symmetry manifest.

\noindent
We start again by  the evolution equations written In terms of the logarithmic variables (\ref{llog}), but now we use different local coordinates. By using as a new basis the eigenvectors
$v^\pm$, we will get a decomposition of the following form
\[y_j=y_+v_j^+ + y_- v_j^- + \sum_{k=1}^N R_k \tau^{(k)}_j\]
\noindent
As $v^+=\bar v^-$, since $y_j$ are real, it holds $y_-= \bar y_+$, $x_j \in {\mathbb R}$ and the term in the sum is the same as in (\ref{decompoy}).
To keep working with real quantities, it is natural to set:
\begin{eqnarray}
y_\pm=(1/{\sqrt 2} )(p\pm iq),\label{coordinates}\\
v^\pm= (1/{\sqrt 2})(u\pm iw).\label{realbasis}
\end{eqnarray}

\noindent
It follows:
\[y_j=pu_j-qw_j+  \sum_{k=1}^N R_k \tau^{(k)}_j\]

\noindent
To make a close comparison with the procedure and the results obtained in Section 2, we write the eigenvalue equation (\ref{eigeq})
In terms of the variables $\epsilon,\eta$, getting:

\[\sum_{k=1}^N(\epsilon_k\eta_j-\epsilon_j\eta_k)v_k=\lambda v_j\]

\noindent
By setting

\[\alpha=\sum_{k=1}^N\epsilon_kv_k,\;\;~\beta=\sum_{k=1}^N\eta_kv_k,\]

\noindent
 we can  write:

\[\eta_j\alpha-\epsilon_j\beta = \lambda v_j.\]

\noindent
Through obvious manipulations  we arrive at the following homogeneous linear system:
\begin{eqnarray}
|\eta|^2\alpha-(\eta,\epsilon)\beta = \lambda \beta\label{eigeneq1}\\
(\epsilon,\eta)\alpha - |\epsilon|^2\beta=\lambda \alpha\label{eigeneq2}
\end{eqnarray}


\noindent
The solution of the secular equation yields the following two complex conjugated eigenvalues:

\begin{equation}\label{eqom}
\lambda_\pm \doteq \pm i\omega = \pm i{\sqrt {|\epsilon|^2|\eta|^2-(\epsilon,\eta)^2}}.
\end{equation}
Notice that $\omega$, i.e. the expression under the square root, is positive because of Cauchy-Schwartz inequality \cite{Kreyszig}, as already noticed in the previous section. Explicitly we have:
\begin{equation}
|\epsilon|^2|\eta|^2-(\epsilon,\eta)^2=-\frac{1}{2}\textrm{Tr}(A^2)=\sum_{k,i>k}(\epsilon_i\epsilon_k)^2(B_i-B_k)^2.
\end{equation}

 \noindent
Once solved for $\alpha,\beta$ the system (\ref{eigeneq1}), (\ref{eigeneq2}),  in terms of the variables $\epsilon,\eta$ we get:

\begin{eqnarray}
u_j=\sqrt{2}\rho\epsilon_j\label{u}\\
w_j=\frac{\sqrt{2}\rho}{\omega}(\epsilon_j(\epsilon,\eta)-\eta_j|\epsilon |^2)\label{v}\\
\rho^2=1/(2|\epsilon|^2)\label{ro}
\end{eqnarray}

\noindent
A straightforward calculation yields the normalization properties:

\begin{equation}
|u|^2=|w|^2=1,~(u,w)=0.\label{norms}
\end{equation}
 
\noindent
Moreover, it turns out that:

\[A_{jk}=\epsilon_k\eta_j-\epsilon_j\eta_k=\omega(u_jw_k-w_ju_k)\]

\noindent
implying that the evolution equations (\ref{llog}) become:
\begin{equation}
\dot p u_j-\dot q w_j =\frac{1}{\sqrt{2}\rho}u_j+\omega (u_j\sum_{k=1}^N \mathcal{C}_kw_k\exp(pu_k-qw_k)-w_j\sum_{k=1}^N\mathcal{C}_k u_k\exp(pu_k-qw_k )
\end{equation}

\noindent
Whence:
\begin{eqnarray}
\dot p =\frac{1}{\sqrt{2}\rho} +\omega \sum_{k=1}^N{\mathcal C}_k w_k\exp(pu_k-qw_k)\label{pdot}\\
\dot q = \omega\sum_{k=1}^N{\mathcal C}_k u_k \exp(pu_k-qw_k)\label{qdot}
\end{eqnarray}

The Hamiltonian nature of equations (\ref{pdot},\ref{qdot}) stems easily from the original Poisson structure (\ref{PB}), that implies

\[\dot q = \omega \frac{\partial {\mathcal H}}{\partial p}\]
\[\dot p =-\omega \frac{\partial {\mathcal H}}{\partial q}\]

The two expressions coincide for the Hamiltonian:
\begin{equation}
{\mathcal H}=  \sum_{k=1}^N{\mathcal C}_k\exp(pu_k-qw_k)-\frac{q}{\sqrt{2}\omega\rho} \label{newHam}
\end{equation}
We conclude this Section by noticing that the variables $(p,q)$ and $(P,Q)$ are related by a linear combination, i.e.
\begin{equation}\label{tra}\left\{
\begin{split}
&p=\frac{1}{\sqrt{2}\rho}P+\sqrt{2}\rho(\epsilon,\eta)Q,\\
&q=\sqrt{2}\rho\omega Q.
\end{split}
\right.\end{equation} 
This transformation is not canonical: it is possible to get a  canonical transformation by the rescaling $(p,q)$ $\to$ $(\sqrt{2}\rho p, \frac{q}{\sqrt{2}\rho\omega})$

 \bigskip
\section{Old and new results: how to understand them?}\label{sec3}
\noindent
In \cite{SRTZ}  we constructed a complete set of first integral in involution for the system (\ref{model}) with the interaction coefficients given by (\ref{matrix}), showing that out of the family
\begin{equation}
e^{-a\sum_i N_i(t)}\prod_{i=1}^N N_i(t)^{c_i}=I_{1,..,N},\label{eqnew}
\end{equation}
\noindent
one can extract $N-1$ independent integrals of motion in involution, with respect to the Poisson brackets (\ref{PB}). In (\ref{eqnew}) $c_k$ are a set of arbitrary constant constrained by the equation $\sum_k c_k \epsilon_k=0$ whereas $a=\sum_{k=1}^N B_k\epsilon_{k}c_k$. Clearly it is always possible to rescale the constant $c_k$ and set $a=1$ so that the constants $c_k$ can be considered the set of equilibrium configurations $z_k$ given in formulae (\ref{coordZ}).

It is  natural to ask  the following question: what is the role of the integrals of motions $\tilde C_k$ found out in \cite{SRTZ}?
As we will show, the answer to this question will be clear if one introduces again the variables $(P,Q)$. First of all, we remind that in term of the logarithmic variables $y_j$
 the generating function of the first integrals  can be written as a single exponential of the quantity:
\[
I (c_1\, \cdots, c_N) =\sum_{k=1}^N \epsilon_kc_k C_k 
\]
\noindent
where
\begin{equation}
C_k = (y_k/\epsilon_k -B_k\sum_{j=1}^N\exp (y_j))\label{newfirst}
\end{equation}

\noindent
In terms of the Poisson bracket (\ref{PB}) we obtain
\begin{equation}\label{inv}
\{C_k,C_l\}= (B_k-B_l)[1-\sum_{j=1}^N\epsilon_jB_j\exp{(y_j)}]
\end{equation}
of course, under the conditions that the matrix $A$  be of the form (\ref{matrixform}).

\noindent
The formula above entails  that the differences $C_k-C_r, C_l-C_r$, $r$ being arbitrary but fixed, $k,l$ running from $1$ to $N$, provide a family of $N-1$ integrals of motion in involution.

\noindent
In terms of the $P,Q$  variables we have (up to an irrelevant additive constant):
\begin{equation}\label{CPQ}
C_k= P+B_k\left(Q -\sum_{j=1}^N {\mathcal C}_j\exp\left(P\xi_j+Q\eta_j\right)\right)=P-B_k{\mathcal H}%
\end{equation}

\noindent
whence it follows
\begin{enumerate}
\item
The quantities $C_k$ are functionally independent, since the Jacobian of the pair $C_k,C_l$ is $\ne 0$, being in fact equal to $(B_k-B_l)\frac{\partial S}{\partial Q}$, where
$S\equiv \sum_j {\mathcal C}_j\exp({P\xi_j+Q\eta_j)}$, but are not first integrals.
\item 
On the other hand.the differences $C_k-C_r$, $C_l-C_r$ ($r$ fixed, $k,l$ running from $1$ to $N$), Poisson commute between themselves and with the Hamiltonian, but are of course functionally 
dependent, among themselves and on the Hamiltonians, as they are simply given by $(B_r-B_k){\mathcal H}$.
\end{enumerate}

\noindent
Our conclusion is that the quantities $\tilde C_k \equiv C_k-C_r$ are genuine independent integrals of motion for the original system  with $N$ degrees of freedom.  However, once the original system is reduced on ${\mathcal I}$ so being transformed into a Hamiltonian system  with one degree of freedom, those quantities are no longer independent and in fact become proportional to the Hamiltonian, which stays as the unique integral motion for the reduced system.

\bigskip

\bigskip

\section{The linearized system}\label{seclin}
It seems not easy to understand under which conditions the dynamical system defined by the Hamiltonian (\ref{h1}) produces closed orbits and periodic motion. In this respect, it may be useful to linearize the original system (\ref{model}) around an equilibrium configuration. We remember that an equilibrium configuration is any vector with elements $z_k$ satisfying the conditions
\begin{equation}\label{eqcon}
\sum_{k=1}^N \epsilon_k z_k=0, \quad \sum_{k=1}^N \eta_k z_k =1
\end{equation} 
\noindent
stemming from the matrix equation:
\[ 
\epsilon_j +\sum_{k=1}^N A_{j,k}z_k = 0, \quad j=1,\ldots,N.
\]
We remember that a coordinate realization of the set of equilibrium configurations is given by equations (\ref{coordZ}). 
We set
\begin{equation}
N_k(t)= z_k+ \delta_k(t) \label{linearizing},
\end{equation}
where the $\delta_k$'s are assumed to be small. At first order we get from (\ref{model})
\begin{equation}
\dot{\delta}_j=z_j\sum_{k=1}^N A_{j,k}\delta_k\doteq \sum_{k=1}^N M_{j,k}\delta_k.
\end{equation}
The matrix with elements $M_{j,k}=z_jA_{j,k}$ defines the local dynamics of the model (\ref{model}). We notice that the matrix $M_{j,k}$ just defined, like the matrix $A_{j,k}$, has rank 2. Its characteristic polynomial is given by
\begin{equation}
\lambda^{N-2}\left(\lambda^2-\frac{1}{2}\textrm{Tr}(M^2)\right)=0,
\end{equation}
where 
\begin{equation}\label{TN}
\textrm{Tr}(M^2)= -\sum_{j,k=1}^N z_jz_k(\epsilon_j\eta_k - \epsilon_k\eta_j)^2
\end{equation}
Since the two eigenvalues different from zero have opposite signs, if they are real the family of equilibrium configurations is unstable.  On the contrary, if the  trace of $M^2$ is negative one has a stable configuration (if the matrix $M$ possesses $N$ independent eigenvectors). From (\ref{TN}) we notice also that it is sufficient, in order to have a negative trace, that the coefficients $z_k$ be all  positive. This in turn implies that at least one of the $\epsilon_k$ is  negative, otherwise equation (\ref{eqcon}) cannot be satisfied. So we expect to find closed orbits in the case when the growth coefficients $\epsilon_k$ are of different signs and to find open orbits when the growth coefficients are all of the same sign.

In the case of periodic motion it is possible to give an interpretation of the equilibrium configurations in terms of the average values of the numerosities. Indeed, let us define the average numerosities as the integral over a period normalized by the length of the period itself, i.e.
\begin{equation}
\overline{N}_j=\frac{1}{T}\int_{0}^T N_j(t)dt.
\end{equation}
Then, from the equations (\ref{model}) we get
\begin{equation}
\ln(N_j(T))-\ln(N_j(0))=0=\epsilon_j+\sum _{k=1}^N A_{j,k}\overline{N}_k,
\end{equation}
showing that indeed the vector with components $\overline{N}_j$ belong to the set of equilibrium configurations.

It might be interesting to consider the linearization of the dynamical system as it appears when written in terms of $P,Q$ variables. 
We start from the equations:

\begin{eqnarray}
\dot P= 1-\sum_{k=1}^N\eta_k\exp(P\epsilon_k+Q\eta_k){\mathcal C}_k\label{eveq3a1}\\
\dot Q = \sum_{k=1}^N\epsilon_k\exp(P\epsilon_k+Q\eta_k){\mathcal C}_k\label{eveq3b1}
\end{eqnarray}

\noindent
An  equilibrium configuration is  a pair ($P^0, Q^0$) satisfying the equations:
\begin{eqnarray}
0= 1-\sum_{k=1}^N\eta_k\exp(P^0 \epsilon_k+ Q^0 \eta_k){\mathcal C}_k \label{equil1}\\
0= \sum_{k=1}^N\epsilon_k\exp(P^0 \epsilon_k+ Q^0 \eta_k){\mathcal C}_k\label{equil2}
\end{eqnarray}

\noindent
So,  the equilibrium configurations are defined as:

\begin{equation}
z_k:=\exp(P^0 \epsilon_k+Q^0 \eta_k){\mathcal C}_k\label{eqnum}
\end{equation}

\noindent
We linearize around the equilibrium configuration, setting:
\[P= P^0 + \delta P,~Q= Q^0+\delta Q,\]
where $\delta P$ and $\delta Q$ are small quantities. From the equations of motion (\ref{eveq3a1}) and (\ref{eveq3b1}) we get, at first order in $\delta P$ and $\delta Q$
\begin{eqnarray}
\delta \dot P = \ell_{1,1}\delta P+\ell_{1,2}\delta Q\label{rodot}\\
\delta \dot Q = \ell_{2,1}\delta P+\ell_{2,2}\delta Q\label{sdot}
\end{eqnarray}
where we set
\[\ell_{2,2}=-\ell_{1,1}=\sum_{k=1}^N z_l\epsilon_k\eta_k,~ \ell_{1,2}=\sum_{k=1}^N z_k \eta_k^2,~\ell_{2,1}=\sum_{k=1}^N z_k\epsilon_k^2.\]
\noindent

\noindent
We rewrite $(\ref{rodot}),(\ref{sdot})$ in matrix form. By denoting
\begin{equation}
\zeta=
\begin{pmatrix}
\delta P\\
\delta Q 
\end{pmatrix}, \quad L=
\begin{pmatrix}
\ell_{1,1}&\ell_{1,2}&\\
\ell_{2,1} & \ell_{2,2}&
\end{pmatrix}
\end{equation}
\noindent
one has:
\begin{equation}
\dot \zeta = L\zeta 	\label{le}
\end{equation}

\noindent
The secular equation associated to the matrix $L$ reads:
\begin{equation}\label{L2}
\lambda^2=-\frac{1}{2}\sum_{jk}z_j z_k(\epsilon_j\eta_k-\epsilon_k\eta_j)^2
\end{equation}
\noindent
Again we can repeat the considerations after equation (\ref{TN}): from equation (\ref{L2}) it follows that  if the equilibrium configuration numerosities $z_j$ have all the same sign (typically, they are all positive) the motion is bounded and periodic. Otherwise there might be open orbits.
Of course (\ref{le})  is  trivially solvable, yielding (with $ \lambda =i\nu)$:

\begin{equation}
\zeta(t)=\exp(Lt)\zeta(0)=[\cos (\nu t)\dblone+i(\sin (\nu t)/\nu) L]\zeta(0)
\end{equation}
where $\dblone$ is the identity matrix.

Finally, in this Section, we would like to remember a result about the compactness of the orbits in the fully nonlinear case given in \cite{SRTZ}. If it is possible to choose the constants $c_k$ in (\ref{eqnew}) to be positive and if the initial conditions $N_k(0)$ are such that the following relation is satisfied:
\begin{equation}
\prod_{k=1}^N \frac{N_k(0)^{c_k}}{e^{aN_k(0)}}<\prod_{k=1}^N \left(\frac{c_k}{ae}\right)^{c_k},
\end{equation}
then the motion occurs on a closed surface isomorphic to the $N$-sphere, explicitly given by
\begin{equation}
\prod_{k=1}^N \frac{N_k(t)^{c_k}}{e^{aN_k(t)}}=\prod_{k=1}^N \frac{N_k(0)^{c_k}}{e^{aN_k(0)}}.
\end{equation}
Clearly, since the constants $c_k$ are constrained to satisfy $\sum_{k}\epsilon_kc_k=0$, it is possible to choose the constants $c_k$ all positive only if the $\epsilon_k$ are not all of the same sign. This result confirms what found with the linearization and gives a sufficient condition to get closed orbits in the nonlinear case.

\section{Graphics and Numerics}\label{sec5}
\begin{figure}[H]
\centering
\includegraphics[scale=0.5]{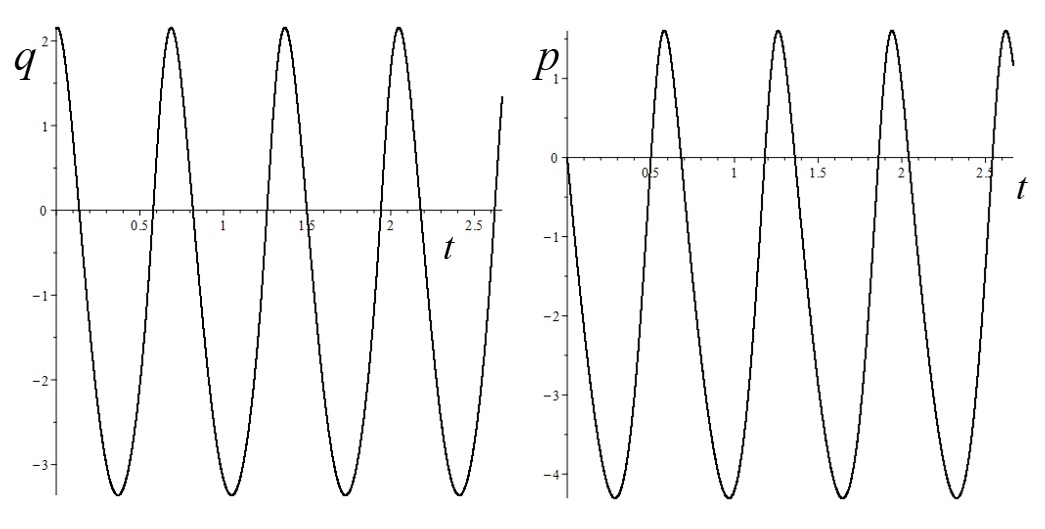}
\caption{Plot of the trajectories of the canonical variables $q$ and $p$ corresponding to the initial conditions $q(0)5\cdot\sqrt{7/38}$ and $p(0)=0$. }
\label{fig1}
\end{figure}

\begin{figure}[H]
\centering
\includegraphics[scale=0.5]{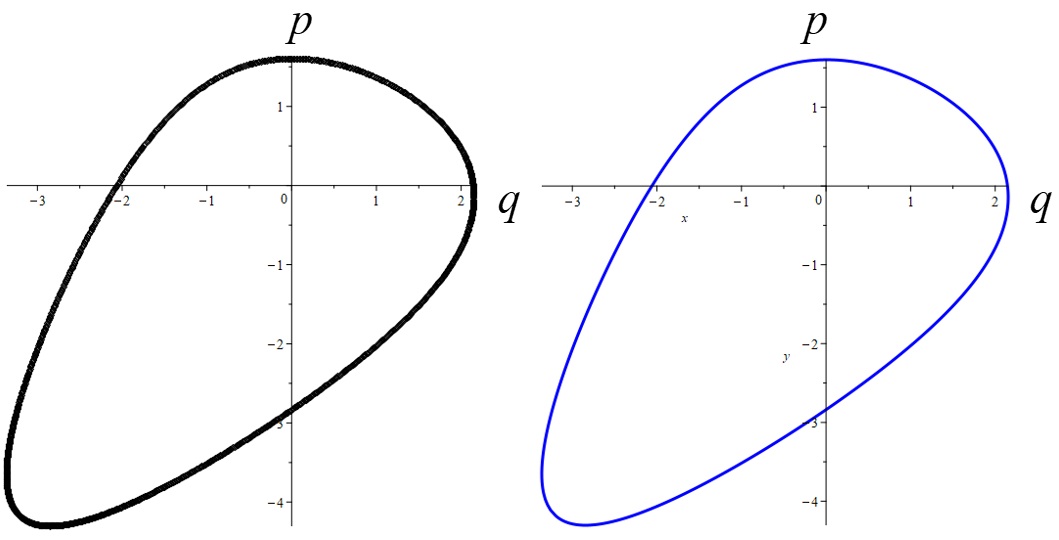}
\caption{Plot of the numerical orbit in the phase space corresponding to the initial conditions $q(0)=5\cdot\sqrt{7/38}$ and $p(0)=0$ (in black, left) and the corresponding plot of the analytical level curve given by $\mathcal{H}(q,p)=(-35 + 76e + 38e^{-1} + 38e^2)/\sqrt{38}$ (blue, right).}
\label{fig2}
\end{figure}

In this section we would like to give some numerical examples of the dynamics defined by the Hamiltonian (\ref{h1}) or (\ref{newHam}). We start with an example of a closed orbit for 4 interacting species. The parameters are fixed in the following way:
\begin{equation}\label{eb}\begin{split}
&\epsilon_1=1, \epsilon_2=-1,\;\epsilon_2=1,\;\epsilon_4=2,\\
&B_1=B_3+3,\; B_2=B_3+1,\; B_4=B_3+2.
\end{split}\end{equation}
The corresponding interaction matrix reads
\begin{equation}A=
\begin{pmatrix}
0 & -2 & 3 & 2\\
2 & 0 & -1 & 2\\
-3 & 1 & 0 & -4\\
-2 & -2 & 4 &0
\end{pmatrix}
\end{equation}
We look firstly at the dynamical system in terms of the canonical variables $(q,p)$ defined by equations (\ref{pdot}) and (\ref{qdot}). The corresponding values of $\omega$ and $\rho$ in (\ref{eqom}) and (\ref{ro}) are respectively given by $\sqrt{38}$ and $1/\sqrt{14}$. The two Casimirs for the set of equations (\ref{llog}) defined by the Kernel of the matrix $A$ are $H_1=4y_2+2y_3+y_4$ and $21y_1-y_2+10y_3-16y_4$. At this point one has to choose the initial conditions. We choose $y_1(0)=1, y_2(0)=2, y_3(0)=-1$ and $y_4(0)=1$ corresponding to the values $q(0)=5\cdot\sqrt{7/38}$ and $p(0)=0$ for the variables $p$ and $q$. The numerical trajectories for the variables $p(t)$ and $q(t)$ are displayed in figure (\ref{fig1}). 
The motion is periodic. Indeed, the corresponding level curve of the Hamiltonian, given by $\mathcal{H}(q,p)=(-35 + 76e + 38e^{-1} + 38e^2)/\sqrt{38}$ is given in figure (\ref{fig2})

\begin{figure}[H]
\centering
\includegraphics[scale=0.5]{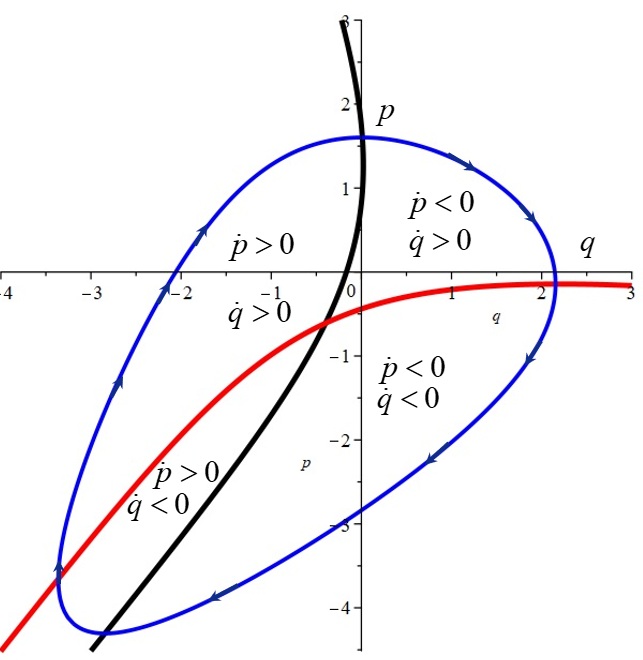}
\caption{Plot of the trajectories of the canonical variables $q$ and $p$ corresponding to the initial conditions $q(0)=5\cdot\sqrt{7/38}$ and $p(0)=0$ together with the curves determined by $\dot{p}=0$ (black) and $\dot{q}=0$ (red). The arrows show the direction of the flow.}
\label{fig5}
\end{figure}

We can plot also the lines $\dot{p}=0$ and $\dot{q}=0$ corresponding to the solutions of the right hand side of equations (\ref{pdot}) and (\ref{qdot}). They divide the plane $(p,q)$ in four regions, each possessing a precise sign of $p$ and $q$. The curves meet at the equilibrium point. The corresponding plots are given in Figure (\ref{fig5})

\begin{figure}[H]
\centering
\includegraphics[scale=0.5]{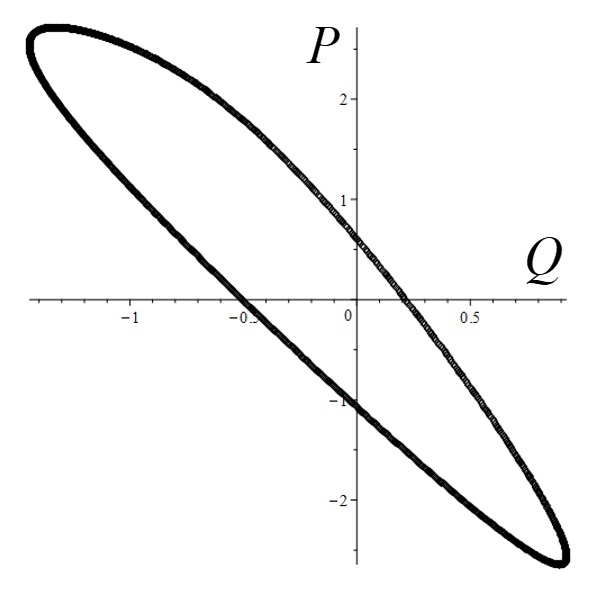}
\caption{Plot of the trajectory in the plane $(Q,P)$ corresponding to the initial conditions $Q(0)=35/38$ and $P(0)=-5/2$.}
\label{fig6}
\end{figure}

We can give also a plot of the previous closed trajectory in the plane $(P,Q)$. We notice, however, that the equations of motion (\ref{eveq3a}-\ref{eveq3b}), and hence the trajectories, depend on all the parameters $B_i$, differently from the equations of motion (\ref{pdot}-\ref{qdot}) that depends only on the differences $B_i-B_j$. So we must also fix the value of $B_3$ in (\ref{eb}). Let us choose $B_3=1$. The values of the initial conditions are $P(0)=-5/2$ and $Q(0)=35/38$ correspondingly to the dame initial conditions for the variables $y_i$, $i=1,...,4$, as before. The plot is given in Figure (\ref{fig6}). The coordinate $Q$ depend only on the differences $B_i-B_j$ whereas, as can be seen from equations (\ref{tra}), $P$ depends linearly on $B_3$, so by changing this value one has a stretching of the closed curve given in Figure (\ref{fig6}) 

\begin{figure}[H]
\centering
\includegraphics[scale=0.5]{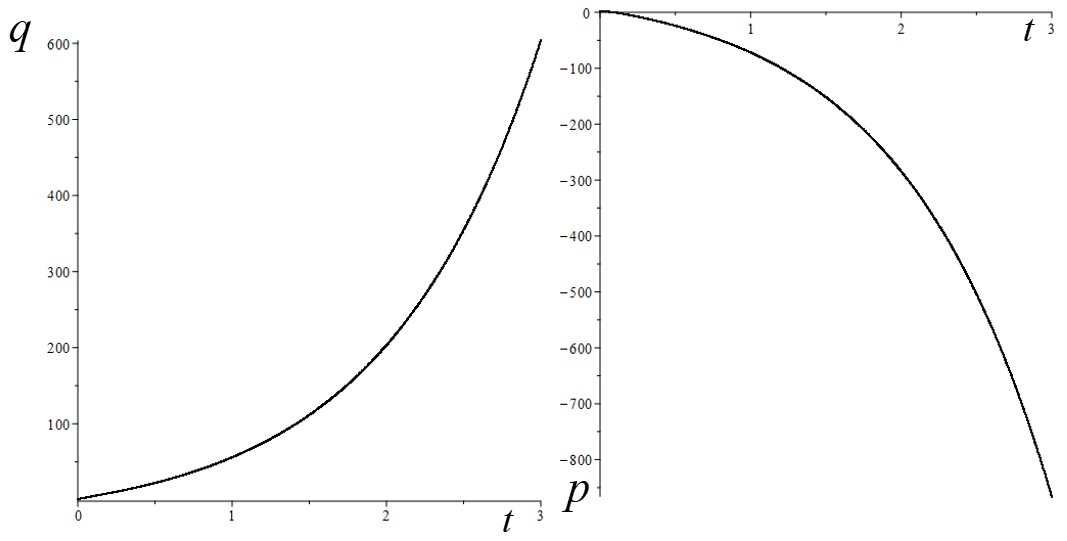}
\caption{Plot of the trajectories of the canonical variables $q$ and $p$ corresponding to the initial conditions$q(0)=15/\sqrt{266}$ and $p(0)=4/\sqrt{7}$ . }
\label{fig3}
\end{figure}

\begin{figure}[H]
\centering
\includegraphics[scale=0.5]{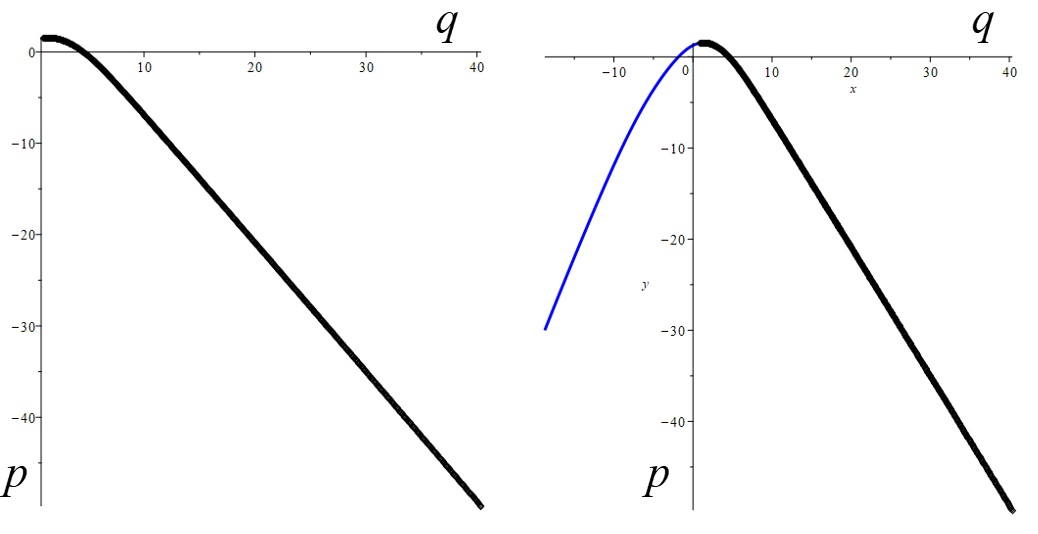}
\caption{Plot of the numerical orbit in the phase space corresponding to the initial conditions $q(0)=15/\sqrt{266}$ and $p(0)=4/\sqrt{7}$ (in black, left) and the corresponding plot of the analytical level curve given by $\mathcal{H}(q,p)=(38e^2+76e-15+38e^{-1})/\sqrt{38}$ (blue, right). The black curve on the right is overlaying the blue one.}
\label{fig4}
\end{figure}

In the next example we flip the value of $\epsilon_2$: from negative to positive. So we set:
\begin{equation}\begin{split}
&\epsilon_1=1, \epsilon_2=1,\;\epsilon_2=1,\;\epsilon_4=2,\\
&B_1=B_3+3,\; B_2=B_3+1,\; B_4=B_3+2.
\end{split}\end{equation}
The corresponding interaction matrix reads
\begin{equation}A=
\begin{pmatrix}
0 & 2 & 3 & 2\\
-2 & 0 & 1 & -2\\
-3 & -1 & 0 & -4\\
-2 & 2 & 4 &0
\end{pmatrix}
\end{equation}
The values of $\rho$ and $\omega$ are the same as before, whereas the two integrals of motion for the set of equations (\ref{llog}) defined by the Kernel of the matrix $A$ are $H_1=-4y_2+2y_3+y_4$ and $21y_1+y_2+10y_3-16y_4$. We fix the initial conditions to be the same as before, i.e. $y_1(0)=1, y_2(0)=2, y_3(0)=-1$ and $y_4(0)=1$ corresponding to the values $q(0)=15/\sqrt{266}$ and $p(0)=4/\sqrt{7}$  for the variables $p$ and $q$. The numerical trajectories for the variables $p(t)$ and $q(t)$ are displayed in figure (\ref{fig3}). 
Now the motion is unbounded. The corresponding level curve of the Hamiltonian, given by $\mathcal{H}(q,p)=(38e^2+76e-15+38e^{-1})/\sqrt{38}$ is given in figure (\ref{fig4})

\noindent The analytical, numerical, and graphic outcomes displayed in this section confirm and consolidate the results derived in (\cite{SRTZ}).

\bigskip
\section{Concluding Remarks}\label{sec6}

We have seen that all the commuting integrals of motion $\tilde C_k\equiv C_k-C_1$, when evaluated on ${\mathcal I}$, are functionally dependent among themselves and on the Hamiltonian. Nevertheless it might be interesting to identify their dynamical meaning, in other words the flow that they generate. Let us define the bracket $\{y_i,\tilde C_k\}$ as $\frac{\partial y_i}{\partial t_k}$, and $\{y_i,{\mathcal H}\}$ as $\frac{\partial y_i}{\partial t_H}$. By looking at equation (\ref{CPQ}) it is clear that one has
\[\frac{\partial y_i}{\partial t_k}=(B_1-B_k)\frac{\partial y_i}{\partial t_H},\]
\noindent
meaning that
\[t_k=t_H/(B_1-B_k).\]
\noindent
The flow generated by the commuting integrals of motion $\tilde{C}_k$ is the same as the one generated by the Hamiltonian, with a properly contracted or dilated  time-scale.

Also, we would like to answer to the following question: do the integrals of motion introduced by Vito Volterra in 1937 (\cite{Volterra}), in his framework with doubled number of variables and canonical Poisson bracket, have any correspondence with those derived by us in our recent paper (\cite{SRTZ})? Let us remember how Volterra doubled the number of coordinates. He introduced the quantity  of life for each species $k$ by defining $q_k = \int_0^t N_k(\tau)d\tau$. The linear momenta, canonically conjugated to the quantities of life, are then defined as \cite{SRTZ}:
\begin{equation} 
p_k = \log \dot q_k  +1 - \frac{1}{2}\sum_{n=1}^N A_{kn}q_n\label{momenta}
\end{equation}
whereas the Hamiltonian $H_V$ of the model is given by:
\begin{equation}
H_V=\sum_{k=1}^N \epsilon_k q_k-\dot{q}_k
\end{equation}
where one has to substitute the value of $\dot{q}_k$ in terms of $p_k$ and the $q$'s from (\ref{momenta}). Then, Volterra shows that the quantities
\begin{eqnarray*}
{H}_r-{H}_l= \frac{p_r-\frac{1}{2}\sum_s A_{r,s}q_s}{\epsilon_r} - \frac{p_l-\frac{1}{2}\sum_s A_{l,s}q_s}{\epsilon_l} 
\end{eqnarray*}
are functionally independent conserved quantities also when the matrix $A$ is the degenerate one leading to complete integrability. By taking also into account equations  (\ref{momenta}), one has:

\begin{equation}
{H}_r-{H}_l=\frac{\log (\dot q_r)+1}{\epsilon_r}-\frac{\log (\dot q_l)+1}{\epsilon_l}-(B_r-B_l)\sum_{s=1}^N\epsilon_s q_s
\end{equation}
The relations among $\mathcal{H}$, $H_V$ and the conserved quantities $C_k-C_n$ and $H_k-H_n$ can be obtained by using formulae (\ref{newfirst}) and (\ref{CPQ}). We can write:
\begin{equation}\label{rel}
H_k-H_n+(B_k-B_n)H_V=C_k-C_n+(B_k-B_n)\mathcal{H}+D_{k,n},
\end{equation}
where $D_{k,n}$ is  a suitable constant depending on the parameters.  Clearly, the conserved quantities $H_k-H_n$ and $H_V$ depend also on the integral of the numerosities, whereas the integrals $C_k-C_n$ and $\mathcal{H}$ depend punctually on the numerosities: a direct relationship between $H_k-H_n$ and $C_k-C_n$ would be not possible, since the integrals of the numerosities must be balanced by the term proportional to $H_V$ on the right hand side of (\ref{rel}). In other words the differences between $H_k-H_n$ and $C_k-C_n$ is a conserved quantity and must be proportional to $H_V$, whereas the difference between $H_V$ and a suitable combination of $H_k-H_n$ is a conserved quantity depending only on the numerosities, not their integral, and then must be proportional to $\mathcal{H}$. These considerations are quantified by equations (\ref{rel}).

Let us summarize the findings of this work: we have shown that the Volterra's integrable system with $N$-species can be reduced to an Hamiltonian system with only one degree of freedom. The corresponding motion can be bounded (and, in fact, periodic), or unbounded. We give a sufficient condition, depending on the parameters of the model and on the initial conditions, to get a bounded motion.  The problem to establish, for a suitable choice of the parameters, how many equilibrium configurations exist for the reduced system in the canonical coordinates $(P,Q)$ or $(p,q)$ could be an interesting question: as a matter of fact we always get just one equilibrium configuration in the plane $(P,Q)$ (see e.g. Figure (\ref{fig5}), where the equilibrium configuration is given by the intersection of the red and black curves). The fact that the bounded motion take place on a hypersurface isomorphic to the $N$-sphere seems to strength the conjecture that the system of equations (\ref{equil1})-(\ref{equil2}) possesses one solution for bounded motion and zero solutions for unbounded motion, depending on the values of the parameters. But we cannot give here a proof. If the conjecture would be true, we can exclude also the existence of more exotic orbits like homoclinic or heteroclinic orbits. 

Finally, other possible, and in our opinion interesting, directions that a forthcoming research could pursue to extend our results are the following: the search for an exact  time discretization of the results derived here, or, equivalently, the search for auto-B{\"a}cklund transformations. A second direction would be the generalization of our results  to a mathematically meaningful $N \to \infty$ limit.

\subsection*{Acknowledgements}

F.Z. wishes to acknowledge the support of Universit\`a degli Studi di Brescia; INFN, Gr. IV - Mathematical Methods in NonLinear Physics and ISNMP - International Society of Nonlinear Mathematical Physics. O.R. and F.Z. wish to acknowledge the support of GNFM-INdAM.

\label{lastpage}
\end{document}